\documentclass[10pt,twocolumn]{article} 
\usepackage{ieee_cs_latex/latex8}
\usepackage{times}

\RequirePackage{ifthen} 

\usepackage{amsbsy}             

\makeatletter
\newcommand{\mathfn}[1]{\mathop {\operator@font #1}\nolimits }
\makeatother

\newcommand{\lmax}{\mathfn{\sf lmax}}
\newcommand{\lmin}{\mathfn{\sf lmin}}
\newcommand{\gr}{\mathfn{\sf partial}}

\newcommand{\dx}{\boldsymbol{\alpha}}
\newcommand{\dy}{\boldsymbol{\beta}}

\newcommand{\ratio}{\mathfn{\sf ratio}}
\newcommand{\local}{\mathfn{\sf local}}
\newcommand{\Global}{\mathfn{\sf global}}

\newcommand{\opt}[1]{#1^*}  

\renewcommand{\v}[1]{{\boldsymbol{#1}}}
\renewcommand{\v}[1]{{\mathbf{#1}}}

\newcommand{\vs}[2]{\v{#1}_{#2}}

\renewcommand{\o}[1]{\opt{\v{#1}}}

\newcommand{\os}[2]{\o{#1}_{#2}}

\newtheorem{lemma}{Lemma}
\newtheorem{corollary}{Corollary}
\newtheorem{definition}{Definition}

\newcommand{\tab}{\hspace*{0.3in}}

\newcommand{\qed}{\vspace{.1em}\noindent\fbox{\rule{
0em}{.1em}\rule{.1em}{0em}}\vspace{1em}}

\newenvironment{proof}{

\noindent{\bf Proof:}\ }{
\hfill \qed

}

\newcommand{\hlinefill}{\parbox{\textwidth}{\hrulefill}}

\newcommand{\algorithmno}[2][]{
  \smallskip
  \noindent
    \ifthenelse{\equal{#1}{}}{\hrulefill\\}{%
      {\parbox{0.25in}{\hrulefill} \sc #1 \hrulefill \smallskip\\}
    }
    \begin{tabular}{l} 
      #2 
    \end{tabular}
    \\\hlinefill
}

\newcommand{\rem}[1]{\hfill \ensuremath{\mbox{\em---#1}}}

\newcommand{\figsize}{\large}
\newcommand{\figspace}{\vspace*{-0.25in}}

\newcommand{\II}{{j}}
\newcommand{\JJ}{{i}}

\begin{document}

\title{
  Sequential and Parallel Algorithms
  for Mixed Packing and Covering
  }
\author{
 Neal E. Young\\
 Akamai Technologies\\
 Cambridge, Massachusetts, USA.\\
 neal@acm.org
 }

\maketitle
\thispagestyle{empty}

\begin{abstract}
  We describe sequential and parallel algorithms that
  approximately solve linear programs with no negative
  coefficients (a.k.a.\ {\em mixed packing and covering
    problems}).
  
  For explicitly given problems, our fastest sequential
  algorithm returns a solution satisfying all constraints
  within a $1\pm\epsilon$ factor in $O(md\log(m)/\epsilon^2)$
  time, where $m$ is the number of constraints and $d$ is the
  maximum number of constraints any variable appears in.
  
  Our parallel algorithm runs in time polylogarithmic in the
  input size times $\epsilon^{-4}$ and uses a total number of
  operations comparable to the sequential algorithm.
  
  The main contribution is that the algorithms solve {\em
    mixed} packing and covering problems (in contrast to pure
  packing or pure covering problems, which have only ``$\le$''
  or only ``$\ge$'' inequalities, but not both) and run in
  time independent of the so-called {\em width} of the
  problem.
\end{abstract}

\Section{Background}
{\em Packing and covering problems} are problems that can be
formulated as linear programs using only non-negative
coefficients and non-negative variables.  Special cases
include pure packing problems, which are of the form $\max
\{c\cdot x : Ax \le b\}$ and pure covering problems, which are
of the form $\min \{c\cdot x : Ax \ge b\}$.

{\em Lagrangian-relaxation} algorithms are based on the following basic
idea.  Given an optimization problem specified as a collection
of constraints, modify the problem by selecting some of the
constraints and replacing them by a continuous ``penalty''
function that, given a partial solution $\v x$, measures how
close $\v x$ is to violating the removed constraints.
Construct a solution iteratively in small steps, making each
choice to maintain the remaining constraints while minimizing
the increase in the penalty function.

While Lagrangian-relaxation algorithms have the disadvantage
of producing only approximately optimal (or approximately
feasible) solutions, the algorithms have the following
potential advantages in comparison to the simplex, interior
point, and ellipsoid methods.  They can be faster, easier to
implement, and easier to parallelize.  They can be
particularly useful for problems that are sparse, or that have
exponentially many variables or constraints (but still have
some polynomial-size representation).

Lagrangian relaxation was one of the first methods proposed
for solving linear programs --- as early as the 1950's, John
von Neumann apparently proposed and analyzed an
$O(m^2n\log(mn)/\epsilon^2)$-time Lagrangian-relaxation
algorithm for solving two-person zero-sum matrix games 
(equivalent to pure packing or covering) \cite{vonNeumann}.  The
algorithm returned a solution with {\em additive} error
$\epsilon$ assuming the matrix was scaled to lie between 0 and
1.  In 1950, Brown and von Neumann also proposed a
system of differential equations that converged to an optimal
solution, with the suggestion that the equations could form
the basis of an algorithm \cite{brown50}.

Subsequent examples include a multicommodity flow algorithm by
Ford and Fulkerson (1958)\nocite{FordF}, Dantzig-Wolfe
decomposition (1960)\nocite{DantzigW}, Benders' decomposition
(1962)\nocite{Benders}, and Held and Karp's lower bound for
the traveling salesman problem (1971)\nocite{HeldK1,HeldK2}.
In 1990, Shahrokhi and Matula \nocite{ShahrokhiM} proved 
polynomial-time convergence rates for a
Lagrangian-relaxation algorithm for multicommodity flow.
This caught the attention of the theoretical computer science
research community, which has since produced a large body of
research on the subject.  Klein et al.\ 
\cite{KleinPST94} and Leighton et al.\ 
\cite{LeightonMPSTT95} (and many others) gave
additional multicommodity flow results.  Plotkin, Shmoys, and
Tardos \cite{PlotkinST94} and Grigoriadis and
Khachiyan
\cite{grigoriadis95exponential,grigoriadis96coordination,grigoriadis96approximate,grigoriadis95sublinear-time}
adapted the techniques to the general class of
packing/covering problems, including mixed packing and
covering problems.  These algorithms' running times
depended linearly on the {\em width} ---
an unbounded function of the input instance.
Relatively complicated techniques were developed to
transform problems so as to reduce their width.
\nocite{karger95adding}
\nocite{bienstock99approximately}

From this body of work we adapt and use the following specific
techniques: the technique of {\em variable-size increments}
(Garg and Konemann \cite{GargK98,konemann}); a way of
partitioning the steps of the Garg/Konemann algorithm into
{\em phases} (Fleischer \cite{Fleischer}); and the idea of
{\em incrementing multiple variables simultaneously} (Luby and
Nisan \cite{LubyN93}).

Variable-sized increments yield algorithms whose running times
are independent of the width of the problem instance,
effectively replacing the width by the number of constraints.
Partitioning into phases reduces the time to implement each
step.  Finally, incrementing multiple variables simultaneously
allows fast parallel algorithms.

Previously, as far as we know, these techniques have only been
applied to pure packing or covering problems, not to mixed
packing and covering problems.  We know of no other
width-independent or parallel algorithms for {\em mixed}
packing and covering.  Our contribution here is to present
such algorithms.

Although Luby and Nisan characterize
their algorithm as solving
``linear programs with non-negative coefficients'',
in fact it applies only to pure packing or pure covering problems
\cite{LubyN93}.

After presenting and analyzing the algorithms,
we conclude with two illustrative examples.

\begin{figure*}[t]\figsize
  \begin{center}
    \parbox{0.90\textwidth}{

      {\bf in:} $\v P, \v C, \epsilon$, arbitrary $\v x$
      \\{\bf out:} 'infeasible' or $\v x$ s.t.\ $\v P\v x \le
      (1+O(\epsilon))N, \v C\v x \ge N$.
      
      {
        1. Let \(N \leftarrow (\max \v P \v x + 2\ln m)/\epsilon\),
        where $m$ is the number of constraints.

        2. Define  \(\gr_\II(\v M, \v x) =
        \sum_\JJ \vs M {\JJ\II} e^{(\v M\v x)_\JJ} / \sum_\JJ e^{(\v M\v x)_\JJ}\).
        \rem{partial derivative of $\lmax$ and $\lmin$}

        3. Define
        \(\ratio_\II(\v x) = \gr_\II(\v P, \v x) / \gr_\II(\v C, -\v x)\).

        4. While $\min \v C \v x < N$ do:

        5.\tab If $\min_\II \ratio_\II(\v x) > 1$ then return
        'infeasible'.
        \rem{see Lemma~\ref{inf}}
        
        6.\tab Delete the $\JJ$th row of $\v C$
        for each $\JJ$ s.t.\ $(\v C \v x)_\JJ \ge N$.
        \rem{constraint deletion for efficiency}
                                
        7a.\hspace{-0.5em}\tab Choose ``increment'' vector $\dx \ge 0$ such that

        7b.\hspace{-0.5em}\tab\tab
        \((\forall \II)\) \(\dx_\II > 0\) only if \(\ratio_\II(\v x) \le 1+\epsilon\)
        \rem{ $\partial \lmax / \partial \vs x j \le (1+\epsilon) \partial  \lmin/ \partial \vs x j$}
        
        7c. \hspace{-0.5em}\tab\tab and $\max\{\max \v C \dx,\max \v P \dx\} = \epsilon$.
        \rem{step size}

        8.\tab Let $\v x \leftarrow \v x + \dx$.
        \rem{do the increment}

        9. Return $\v x$.
        }
      }
    \caption{Generic algorithm.
      Implementable in $O(m\log(m)/\epsilon^2)$ linear-time iterations.
}
    \label{generic}
  \end{center}
  \figspace
\end{figure*}

\Section{Mixed Packing and Covering}
We consider problems in the following form:

\noindent{\sc Approximate Mixed Packing and Covering:}
{
Given
non-negative matrices $\v P,\v C$,
vectors $\v p,\v c$ and $\epsilon \in (0,1)$,
find an {\em approximately feasible} vector
$\v x \ge \v 0$ (s.t.\ 
$\v P \v x \le (1+\epsilon)\v p$ and $\v C \v x \ge \v c$)
or a proof that no vector $\v x$ is feasible (i.e., satisfies $\v x \ge \v 0$,
$\v P \v x \le \v p$ and $\v C \v x \ge \v c$).
}

\smallskip
In Section~\ref{reduce} we describe how to reduce
the optimization version
\(\min\{\lambda : \v P \v x \le \lambda\, \v p, \v C \v x \ge \v c\}\)
to the above form.

For notational simplicity,
we assume throughout that each coordinate of $\v p$ and $\v c$ is some
constant $N$.
This is without loss of generality
by the reduction
$\vs {P} {\JJ\II}' = \vs P {\JJ\II} N/\vs p \JJ$
and $\vs {C} {\JJ\II}' = \vs C {\JJ\II} N/\vs c \JJ$
(after removing any constraints
of the form $({\v P \v x})_\JJ \le 0$
--- which only force to zero each $\vs x \II$ such that $\vs P {\JJ\II} > 0$ ---
or the form $({\v C \v x})_\JJ \ge 0$
--- which do not constrain $\v x$ at all).
A vector $\v x$ is {\em feasible} if
$\max \v P \v x \le N \le \min \v C \v x$.

\smallskip
All of the algorithms in this paper
are specializations of
the generic algorithm in Fig.~\ref{generic}.
The algorithm starts with an infeasible vector $\v x$
and adds to $\v x$ in small increments until $\v x$ becomes
approximately feasible, that is, until
$\min \v C \v x  \ge N$ and $\max \v P\v x \le (1+O(\epsilon))N$.
Instead of working with the $\max$ and $\min$ functions,
the algorithm works to achieve a stronger condition:
$\lmin \v C \v x \v  \ge N$ and $\lmax \v P\v x \le (1+O(\epsilon))N$,
where $\lmax$ and $\lmin$ are ``smooth''
functions that approximate $\max$ and $\min$:
\begin{definition}
  For real values $\v y=(\vs y 1, \vs y 2,\ldots,\vs y m)$, define
  \[
  \lmax \,\v y \, = \, \ln {\textstyle \sum_\JJ e^{\vs y \JJ}}
  \mbox{~~~~and~~~~}
  \lmin \,\v y \, = \, -\ln {\textstyle \sum_\JJ e^{- \vs y \JJ}}.
  \]
  Note \(\max \v y \le \lmax \v y\) and \(\min \v y \ge \lmin \v y.\)
\end{definition}

Recall  that, for any continuous function $f(\v x)$,
increasing $\vs x \II$ by $\delta$ increases $f$ by
approximately $\delta$ times the partial derivative of $f$
with respect to $\vs x \II$.
In lines~2 and~3 of the algorithm, $\gr_\II(\v P, \v x)$
and $\gr_\II(\v C, -\v x)$ are, respectively,
the partial derivatives of
$\lmax(\v P \v x)$ and $\lmin(\v C\v x)$
with respect to $\vs x \II$.
Thus, the condition in line~7b says that a variable
$\vs x \II$ may be increased {\em only if
doing so increases $\lmax(\v P \v x)$
by at most $1+O(\epsilon)$ times
as much as it increases $\lmin(\v C\v x)$}.
We say ``$1+O(\epsilon)$'' instead of $1+\epsilon$
because the partial derivatives only approximate
the actual increases,
and in fact the condition on line~7c is necessary to ensure
that the change in $\v x$ is small enough
so that the partial derivatives do give good approximations.

Because each step increases $\lmax(\v P \v x)$
by at most $1+O(\epsilon)$ times
as much as it increases $\lmin(\v C\v x)$,
the ratio of the two quantities tends to $1+O(\epsilon)$ (or less).
Thus, the algorithm drives $\v x$ to approximate feasibility.

Line~5 ensures that there is a $j$ meeting the condition of
line~7b.  Why is line~5 okay?  Briefly, because the gradients of $\lmax$ and
$\lmin$ have 1-norm equal to 1 (that is, the sum of their
partial derivatives equals 1), one can show that, for {\em
  any} $\v x$ and any feasible $\o x$, the dot product of $\o
x$ with the gradient of $\lmax(\v P \v x)$ is at most $N$,
while the dot product of $\o x$ with the gradient of $\lmin(\v
C \v x)$ is at least $N$.  Since $\o x \ge 0$, this means at
least one partial derivative of $\lmin(\v C \v x)$ is as large
as the corresponding partial derivative of $\lmax(\v P \v x)$.

In the remainder of this section
we give the complete analysis of the performance guarantee.

Let $\gr_\JJ'(\v y) = e^{\vs y \JJ} / \sum_\JJ e^{\vs y \JJ}$,
so that $\gr_\JJ'(\v y)$ is the partial derivative of $\lmax(\v y)$ with
respect to $\vs y \JJ$
and $\gr_\JJ'(-\v y)$ is the partial derivative of $\lmin(\v y)$ with
respect to $\vs y \JJ$.
We will use the following ``chain rule'': for any $\v M,\v x,\dx$,
\begin{equation}\textstyle
  \sum_\JJ (\v M\dx)_\JJ \gr_\JJ'(\v M\v x) =
  \sum_\II  \dx_\II \gr_\II(\v M, \v x)~~~ \label{chainRule}
\end{equation}
We start with a utility lemma.
\begin{lemma}[smoothness of $\lmin$ and $\lmax$]\label{plugh}
  For all $\v y, \dy \ge 0$, if $0 \le \dy_\JJ \le \epsilon \le 1$
  then
  \[\textstyle
  \lmax(\v y + \dy)
  \le \lmax(\v y) + (1+\epsilon)\sum_\JJ \dy_\JJ \gr_\JJ'(\v y)\]
  and
  \[\textstyle
  \lmin(\v y + \dy)
  \ge \lmin(\v y) + (1-\epsilon/2)\sum_\JJ \dy_\JJ \gr_\JJ'(-\v y).\]
\end{lemma}
\begin{proof}
  Using the standard sorts of inequalities
  that underlie Chernoff bounds, namely
  $\ln(1+z) \le z$ (for all $z$)
  and
  $e^\beta - 1 \le (1+\epsilon)\beta$ 
  (for $0\le \beta \le \epsilon \le 1$):
  \begin{eqnarray*}
    \lefteqn{\lmax(\v y + \dy) - \lmax(\v y)}
    \\ & = & \textstyle
    \ln[\sum_\JJ e^{\vs y \JJ+\dy_\JJ} / \sum_\JJ e^{\vs y \JJ}] \\
    & = &\textstyle
    \ln[1 + \sum_\JJ (e^{\dy_\JJ}-1) e^{\vs y \JJ} / \sum_\JJ e^{\vs y \JJ}] \\
    & \le &\textstyle
    \sum_\JJ (e^{\dy_\JJ}-1) e^{\vs y \JJ} / \sum_\JJ e^{\vs y \JJ} \\
    & \le &\textstyle
    (1+\epsilon) \sum_\JJ \dy_\JJ \gr'_\JJ(\v y).
  \end{eqnarray*}
  This proves the first inequality in the statement of the lemma.
  The second inequality follows by an analogous chain of inequalities,
  using
  $1-e^{-\beta} \ge (1-\epsilon/2)\beta$ 
  (for $0\le \beta \le \epsilon \le 1$).
\end{proof}

Because of this smoothness, if the increment is small enough
(i.e.\ the``step-size'' condition on line~7c of the algorithm is met)
the partial derivatives approximate the changes in $\lmin$ and $\lmax$
well --- within a $1\pm\epsilon$ factor:
\begin{lemma}\label{inc}
  In each increment, the increase of $\lmax \v P \v x$
  is at most $\frac{(1+\epsilon)^2}{1-\epsilon/2}$ times 
  that of $\lmin \v C \v x$.
\end{lemma}
\begin{proof}
  When the generic algorithm increments $\v x$ by $\dx$,
  the vector $\dx$ meets the following
  conditions:
  \begin{enumerate}
  \item \((\forall \II) ~\dx_\II > 0 \rightarrow \ratio_\II(\v x) \le 1+\epsilon\);
  \item $\max\{\max \v C \dx,\max \v P \dx\} \le \epsilon$.
  \end{enumerate}
  Adding $\dx$ to $\v x$ adds $\v P\dx$ to  $\v P\v x$.
  From Condition 2 above and Lemma~\ref{plugh},
  it follows that $\lmax \v P\v x$ increases by at most
  $(1+\epsilon)\sum_\JJ (\v P\dx)_\JJ \gr_\JJ'(\v P\v x)$.
  By the chain rule~(\ref{chainRule}), this equals
  $(1+\epsilon)\sum_\II \dx_\II \gr_\II(\v P, \v x).$

  Similarly,  $\lmin \v C\v x$ increases by at least
  \((1-\epsilon/2)\sum_\II \dx_\II \gr_\II(\v C, -\v x).\)
    Since $\dx_\II > 0$ only if
    $\gr_\II(\v P, \v x)  \le (1+\epsilon) \gr_\II(\v C, -\v x)$
    (Condition~1 above), Lemma~\ref{inc} follows.
\end{proof}

Next we show that if the problem instance is feasible there
always exists a choice of $\dx$ meeting the conditions on
lines~7b and~7c of the algorithm.  This is necessary for the
algorithm to be well-defined.

\begin{lemma}\label{inf}
  If the problem instance is feasible,
  then $\forall \v x\, \exists \II : \ratio_\II(\v x) \le 1$.
\end{lemma}
\begin{proof}
  Let $\v x$ be arbitrary and
  let $\o x$ be a feasible solution.
  By the chain rule~(\ref{chainRule}),
  \[\textstyle
  \sum_\II \os x \II \gr_\II(\v P, \v x)
  \,=\,
  \sum_\JJ (\v P\o x)_\JJ \gr_\JJ'(\v P\v x).\]
  Since $(\v P\o x)_\JJ \le N$, and 
  $\sum_{\JJ}\gr_\JJ'(\v P\v x) = 1$,
  the quantity above is at most $N$.

  Likewise, \(\sum_\II \os x \II \gr_\II(\v C, -\v x) \ge N\).
  
  Since $\o x \ge 0$, there must be some $\II$
  such that $\gr_\II(\v C, -\v x) \ge \gr_\II(\v P, \v x)$.
\end{proof}
Lemma~\ref{inc} means that the condition on line~7b
can be met, and clearly by scaling the condition on line~7c can
also be met.

\begin{figure*}[t]\figsize
  \begin{center}
    \parbox{0.90\textwidth}{

      {\bf in:} $\v P, \v C, \epsilon$, arbitrary $\v x$
      \\{\bf out:} 'infeasible' or $\v x$ s.t.\ $\v P\v x \le
      (1+O(\epsilon))N, \v C\v x \ge N$.

      {
        1. Let \(N \leftarrow (\max \v P \v x + 2\ln m)/\epsilon\),
        where $m$ is the number of constraints.

        2. Define \(\local_\II(\v x) = 
        \sum_\JJ \vs P {\JJ\II} e^{(\v P\v x)_\JJ}
        / \sum_\JJ \vs C {\JJ\II} e^{-(\v C\v x)_\JJ}\).
        \rem{terms of $\ratio_\II(\v x)$ that depend on $\II$}

        3. Define \(\Global(\v x) = 
        \sum_\JJ e^{(\v P\v x)_\JJ} / \sum_\JJ e^{-(\v C\v x)_\JJ}\).
        \rem{so $\ratio_\II(\v x) = \local_\II(\v x)/\Global(\v x)$}

        4. While $\min \v C \v x < N$ do:

        5a.\hspace{-0.5em}\tab If $g$ is not yet set or
        \(\min_\II \local_\II(\v x)/g > 1+\epsilon\) then

        5b.\hspace{-0.5em}\tab\tab let $g \leftarrow
        \Global(\v x)$, and
        \rem{start new phase}

        5c.\hspace{-0.5em}\tab\tab if $\min_\II \local_\II(\v x)/g > 1$ then return
        'infeasible'.

        6.\tab Delete the $\JJ$th row of $\v C$
        for each $\JJ$ s.t.\ $(\v C \v x)_\JJ \ge N$.

        7a.\hspace{-0.5em}\tab Choose ``increment'' vector $\dx \ge 0$ such that

        7b.\hspace{-0.5em}\tab\tab \((\forall \II)\) $\dx_\II > 0$ only if
        \(\local_\II(\v x)/g \le 1+\epsilon\)
        \rem{stronger cond'n than in generic alg.}
        
        7c.\hspace{-0.5em}\tab\tab and $\max\{\max \v C \dx,\max \v P \dx\} = \epsilon$.

        8.\tab Let $\v x \leftarrow \v x + \dx$.

        9. Return $\v x$.
        }
      }
    \caption{Algorithm with phases.  Implementable in $O(md \log(m)/\epsilon^2)$ operations,
      where $d$ is the maximum number of constraints any variable
      appears in.  
      }
    \label{phase}
  \end{center}
  \figspace
\end{figure*}

\begin{figure*}[t]\figsize
  \begin{center}
    \parbox{0.90\textwidth}{

      {\bf in:} $\v P, \v C, \epsilon$
      \\{\bf out:} 'infeasible' or $\v x$ s.t.\ $\v P\v x \le
      (1+O(\epsilon))N, \v C\v x \ge N$.

      {
        0. Let $\vs x \II = \min_\JJ 1/(n\vs P {\JJ\II})$ for
        each $\II$, where $n$ is the \# of var's.
        \rem{$\v x$ initialized, not given}

        1. Let \(N \leftarrow (\max \v P \v x + 2\ln m)/\epsilon\),
        where $m$ is the \# of constraints.

        2. Define \(\local_\II(\v x) = 
        \sum_\JJ \vs P {\JJ\II} e^{(\v P\v x)_\JJ}
        / \sum_\JJ \vs C {\JJ\II} e^{-(\v C\v x)_\JJ}\).
          
        3. Define \(\Global(\v x) = 
        \sum_\JJ e^{(\v P\v x)_\JJ} / \sum_\JJ e^{-(\v C\v x)_\JJ}\).

        4. While $\min \v C \v x < N$ do:

        5a.\hspace{-0.5em}\tab If $g$ is not yet set or
        \(\min_\II \local_\II(\v x)/g > 1+\epsilon\) then

        5b.\hspace{-0.5em}\tab\tab let $g \leftarrow
        \Global(\v x)$, and

        5c.\hspace{-0.5em}\tab\tab if $\min_\II \local_\II(\v x)/g > 1$ then return
        'infeasible'.

        6.\tab Delete the $\JJ$th row of $\v C$
        for each $\JJ$ s.t.\ $(\v C \v x)_\JJ \ge N$.

        7a.\hspace{-0.5em}\tab  Choose ``increment'' vector $\dx \ge 0$ such that
        for some $\delta>0$
        \rem{ incr.\ {\em all} allowed $\vs x j$'s,}

        7b.\hspace{-0.5em}\tab\tab \((\forall \II)\) $\dx_\II = \vs x \II /\delta$ if
        \(\local_\II(\v x)/g \le 1+\epsilon\),
        else $\dx_\II = 0$,
        \rem{ prop.\ to current value}
        
        7c.\hspace{-0.5em}\tab\tab and $\max\{\max \v C \dx,\max \v P \dx\} = \epsilon$.

        8.\tab Let $\v x \leftarrow \v x + \dx$.

        9. Return $\v x$.
        }
      }
    \caption{Parallel algorithm.
      Implementable in parallel time polylogarithmic in input size
      times $\epsilon^{-4}$.
      }
    \label{parallel}
  \end{center}
  \figspace
\end{figure*}

From Lemma~\ref{inc}, the basic
performance guarantee follows easily.
\begin{lemma}\label{performance}
  If the problem instance is feasible, 
  the generic algorithm returns an approximately feasible solution.
  Given an initial $\v x$, the algorithm makes 
  $O(m(\max \v P\v x + \log m)/\epsilon^2)$ increments.
\end{lemma}
\begin{proof}
  First we prove the performance guarantee.
  Define
  \(\Phi \,=\, \lmax \v P\v x -\frac{(1+\epsilon)^2}{1-\epsilon/2}\lmin \v C \v x.\)

  Before the first increment
  \(\Phi \le \ln (m \, e^{\max \v P\v x}) + (1+O(\epsilon))\ln m
  < O(N\epsilon)\).
  By Lemma~\ref{inc}, no increment operation increases $\Phi$.
  Deleting a covering constraint increases $\lmin \v C \v x$
  and therefore only decreases $\Phi$.
  Thus, \(\Phi \le O(\epsilon N)\) throughout the course of the algorithm
  and, just before the last increment
  (when $\lmin \v C\v x \le \min \v C\v x < N$),
  \begin{eqnarray*}
    \lmax \v P\v x & \le &  O(\epsilon N) + (1+O(\epsilon))\lmin \v C\v x
    \\ & \le & (1+O(\epsilon))N.
  \end{eqnarray*}
  With the last increment,
  $\max \v P\v x$ increases by at most $\epsilon$,
  so at termination, $\max \v P\v x \le (1+O(\epsilon))N$
  while $\min \v C\v x \ge N$.

  Next we bound the number of increments.
  Let $m_c$ and $m_p$ be the number of rows of $\v C$
  and $\v P$, respectively, so that $m_c+m_p = m$.
  Define
  \(\Psi = \sum_\JJ (\v P\v x)_\JJ + \sum_\JJ ((\v C\v x)_\JJ - N-\epsilon)\).
  It is initially at least $-m_c(N+\epsilon)$,
  and finally at most $m_p(N+\epsilon)$.
  By the ``step-size'' condition in line~7c,
  each increment increases $\Psi$ by at least $\epsilon$.
  Because of the constraint-deletion operations in line~6,
  \((\v C\v x)_\JJ < N\) before each increment
  and so 
  \((\v C\v x)_\JJ < N + \epsilon\) after each increment
  (for each row $\JJ$ remaining in $\v C$).
  Thus, each
  constraint deletion increases $\Psi$.
  Thus, the number of increments
  is at most $m(N+\epsilon)/\epsilon$,
  which gives the desired bound by the definition of $N$.
\end{proof}

To specify a particular implementation of the algorithm,
we need to specify how the initial $\v x$ is chosen
and how the increment $\dx$ is chosen in each iteration.
Here is one straightforward implementation:
initialize $\v x$ to $\v 0$,
and with each increment choose $\dx$ to be a vector
where $\dx_\II = 0$ for all $\II$ except for a single $\II$
such that
\(\gr_\II(\v P, \v x) \le (1+\epsilon)\gr_\II(\v C,-\v x)\).
The value of that $\dx_\II$ is determined by the step-size condition.
This still leaves some flexibility.
For example, one can choose $\II$ so as to minimize
\(\gr_\II(\v P, \v x)  - \gr_\II(\v C,-\v x)\)
or to minimize (within $1+\epsilon$) 
\(\gr_\II(\v P, \v x) / \gr_\II(\v C,-\v x)\).
Clearly, if the problem instance is given explicitly,
then either of these choices can be implemented
in time linear in the number of non-zero entries in the
matrices.  This gives the following corollary:
\begin{corollary}\label{generic_analysis}
  The generic algorithm can implemented to approximately solve
  any explicitly given problem instance
  in $O(m\log (m)/\epsilon^2)$ linear-time iterations, where
  $m$ is the number of constraints.
\end{corollary}

\Section{Algorithm with Phases}
This algorithm specializes the generic algorithm.
In order to speed the computation of the key function
$\ratio_\II(\v x)$, we break it into two components
as $\ratio_\II(\v x) = \local_\II(\v x) / \Global(\v x)$,
where $\local_\II$ captures the terms that depend on $\II$.
As $\v x $ changes, we recompute $\Global(\v x)$ only
occasionally --- at the start of each phase (i.e., iteration
of the outer loop). The algorithm is shown in Fig.~\ref{phase}.
First we discuss how this is a particular implementation
of the generic algorithm, and then we 
prove a stronger time bound.

\begin{lemma}
  The algorithm with phases is a specialization
  of the generic algorithm.
\end{lemma}
\begin{proof}
  We argue that any increment the algorithm
  does is also an allowable increment for the generic algorithm.
  Since
  \(\local_\II(\v x) / \Global(\v x) = \ratio_\II(\v x)\)
  and $\local_\II(\v x)$ and $\Global(\v x)$
  only increase as the algorithm proceeds,
  at all times
  \(\local_\II(\v x) / g \ge \ratio_\II(\v x)\) .
  Thus, any $\dx$ meeting the conditions of the
  algorithm with phases also meets the conditions
  of the generic algorithm.
\end{proof}
This and Lemma~\ref{performance} imply the performance guarantee:
\begin{corollary}
  The algorithm with phases returns an approximately feasible solution.
  Given an initial $\v x$, the algorithm makes
  at most $O(m(\max \v P\v x + \log m)/\epsilon^2)$ increments.
\end{corollary}
Now here is the stronger time bound:

\begin{lemma}
  Given an initial $\v x$,
  the algorithm with phases uses
  $O((\max \v P\v x + \log m)/\epsilon^2)$
  phases.
\end{lemma}
\begin{proof}
  We claim that $\Global(\v x)$ increases
  by at least a $1+\epsilon$ factor each phase.
  By inspection, $\Global(\v x)$ is initially
  at least $1/m$ and finally
  at most $m e^{O(N)} / e^{-O(N)}$,
  so the result will follow.

  To see the claim, note that at the end of a phase,
  $\local_\II(\v x)/g > 1+\epsilon$ for all $\II$,
  i.e.\ $g < \min_\II \local_\II(\v x)/(1+\epsilon)$.
  But, by Lemma~\ref{inf}, at the start of the next phase,
  the same $\v x$ and the next $g$ satisfy
  $\local_\II(\v x)/g = \ratio_\II(\v x) \le 1$ for some $\II$,
  i.e.\ $g \ge \min_\II \local_\II(\v x)$.
\end{proof}

Consider the following
``round-robin'' implementation of the algorithm.
Start with $\v x = \v 0$.
Implement each phase by cycling through the indices $\II$ once.
For each $\II$, as long as \(\local_\II(\v x) / g  \le 1+\epsilon\),
repeatedly increment $\v x$ by
the vector $\dx$ that has
all coordinates 0 except $\dx_\II$,
whose value is determined by the step size.

Maintain the values $(\v P\v x)_\JJ$ and  $(\v C\v x)_\JJ$ for every $\JJ$.
After a variable $\vs x \II$ is incremented,
the only values that change are those
where $\vs P {\JJ\II}$ or $\vs C {\JJ\II}$ are non-zero.
So maintaining these values requires $O(d)$ time,
where $d\le m$ is the maximum number constraints
any variable appears in.

With these values in hand,
the condition  \(\local_\II(\v x) / g  \le 1+\epsilon\)
can be checked in $O(d)$ time.
Since the number of increments is $O(m\log(m)/\epsilon^2)$,
the total time to do increments is  $O(m d \log(m)/\epsilon^2)$.
Other than increments,
each of the $O(\log(m) / \epsilon^2)$ phases requires $O(m d)$ time,
so we have the following corollary:
\begin{corollary}\label{phase_analysis}
  The algorithm with phases can be implemented to approximately solve
  any explicitly given problem instance
  using $O(md \log(m)/\epsilon^2)$ operations,
  where $d$ is the maximum number of constraints any variable
  appears in.
\end{corollary}
Note that this is an improvement on
Corollary~\ref{generic_analysis}
by a factor equal to the number of non-zero entries in the
matrix, divided by $d$ (this factor can be as large as the
number of variables).

\Section{Parallel Algorithm}
Next we further specialize the algorithm to achieve an
efficient parallel implementation.  The algorithm is shown in
Fig.~\ref{parallel}.  The idea is that we start with each
variable having a small but positive value.  Then, in each
increment step, we increase {\em all} allowed variables
(line~7b), each proportionally to its current value.  This
method allows us to give a polylogarithmic bound on the number of
iterations per phase.

\begin{lemma}
  The parallel algorithm is a specialization
  of the algorithm with phases, starting with $\max \v P \v x \le 1$.
  The parallel algorithm makes
  $O(\log(m) \log(n \log(m)/\epsilon)/\epsilon^2)$
  increments per phase.
\end{lemma}
\begin{proof}
  The first claim is true by inspection and the initial choice of $\v x$.

  It remains to bound the number of increments per phase.
  First, we claim that in each increment
  $\delta = \Omega(N/\epsilon)$.
  This is simply because by the choice of $\dx$,
  for some $\JJ$,
  $(\v C \v x)_\JJ/\delta$ or $(\v P \v x)_\JJ/\delta$
  is at least $\epsilon$,
  but $(\v C \v x)_\JJ$ and $(\v P \v x)_\JJ$ are $O(N)$
  throughout the algorithm.
  Thus, for each $\II$, each increment that increases $\vs x \II$
  increases it by at least a $1+\Omega(\epsilon/N)$ factor.
  Since $\vs x \II$ is initially $\min_\JJ 1/n\vs P {\JJ\II}$
  and finally at most $\min_\JJ N/\vs P {\JJ\II}$,
  it follows that
  at most $O(N \log(N n)/\epsilon)$ increments
  increase $\vs x \II$.

  Finally, in each phase, the last increment of the phase
  increases some $\vs x \II$.  In fact {\em each} increment in the phase
  must have increased that $\vs x \II$, since $\local_\II(\v x)$
  only increased during the phase.  Thus, the number of
  increments in the phase is $O(N\log(N n)/\epsilon)$.
\end{proof}

Since each increment can be implemented in parallel in
polylogarithmic time, and the number of increments
is bounded by the number of phases times
the number of increments per phase.
We have the following corollary.
\begin{corollary}
  The parallel algorithm can be implemented to approximately solve
  any explicitly given problem instance
  in parallel time polylogarithmic in the input size
  times $1/\epsilon^4$.
\end{corollary}

\Section{Reducing Optimization to Feasibility}\label{reduce}

Given a problem instance $\v P, \v p, \v C, \v c$ and $\epsilon>0$,
let
\(\lambda^* = 
\min\{\lambda : (\exists \v x)~ \v P \v x \le \lambda\, \v p, \v C \v x \ge \v c\}\).
In this section we describe how to use the algorithms in this
paper to approximately solve this optimization problem ---
that is, to compute a feasible solution $(\lambda,\v x)$ such that
$\lambda^* \le \lambda \le (1+\epsilon)\lambda^*$.

We reduce the optimization problem to a sequence of approximate
feasibility subproblems.  Each subproblem requires
$\epsilon'$-approximately solving
\(\exists ? \v x : \v P \v x \le \lambda'\, \v p, \v C \v x \ge \v c\)
for a particular $\epsilon'$ and $\lambda'$.
We can solve such a subproblem using any of the algorithms in this paper.

\begin{lemma}
  The approximate optimization problem reduces
  to a sequence of approximate feasibility subproblems:
  $O(\log \log m)$ subproblems with $\epsilon' = 1/2$
  and $O(\log 1/\epsilon)$ subproblems
  where the $i$th-to-last subproblem has
  $\epsilon' = \Omega(\epsilon (\frac{4}{3})^i)$.
\end{lemma}

Note that if we solve the subproblems using any algorithm
whose time depends at least linearly on $1/\epsilon$ (such as
the ones in this paper), then the total time to solve the
second set of subproblems (those with $\epsilon' < 1/2$)
dominated by the time used to solve the last such subproblem.

The remainder of this section contains the proof.
The basic idea is to use binary search for $\lambda^*$, solving
a feasibility problem at each step to bound $\lambda^*$.
This lemma gives starting upper and lower bounds:
\begin{lemma}
  Let \(\lambda = \sum_\JJ \min_\II \sum_{\JJ'}  (\vs P {\JJ'\II} /\vs p {\JJ'}) / (\vs C {\JJ\II} /\vs c \JJ)\).  
  Then $\lambda^* \le \lambda \le m^2 \lambda^*$.
\end{lemma}
\begin{proof}
  Recall
  \(\lambda^* = \min\{\lambda : (\exists \v x)~ \v P \v x \le \lambda\, \v p, \v C \v x \ge \v c\}\).
  For each $\JJ=1,\ldots,m_c$,
  consider the following relaxation:
  \[\textstyle\lambda_\JJ^* = \min\{\lambda : (\exists\v x)~ 
  \sum_{\JJ'} (\v P \v x)_{\JJ'}/\vs p {\JJ'} \le m_p\lambda,
  (\v C \v x)_\JJ \ge \vs c \JJ
  \}.\]
  That is, only {\em one} specified covering constraint,
  and the {\em sum} of the packing constraints, need to hold.
  Clearly $\lambda_\JJ^* \le \lambda^*$.
  Furthermore the optimal solution $\v z(\JJ)$ to the $\JJ$th relaxed problem
  is given by finding $\II$ that minimizes
  $\sum_{\JJ'}  (\vs P {\JJ\II'} /\vs p {\JJ'}) / (\vs C {\JJ\II} /\vs c \JJ)$
  and setting all coordinates of $\v z(\JJ)$ to zero
  except $\vs z \II(\JJ) = 1/(\vs C {\JJ\II} / \vs c \JJ)$.
  Correspondingly
  \(\lambda^*_\JJ  =
  \min_\II \sum_{\JJ'}  (\vs P {\JJ\II'} /\vs p {\JJ'}) / (\vs C {\JJ\II} /\vs c \JJ)\).

  To get an $m^2$-approximate solution to the original problem,
  take $\v x = \sum_\JJ \v z(\JJ)$
  and $\lambda = \sum_\JJ m_p \lambda_\JJ^*$.
  The pair $(\v x, \lambda)$
  is a feasible solution to the original problem
  (each covering constraint is met the
  contribution of the corresponding $z(i)$,
  while each packing constraint is met
  because $\sum_{\JJ'}  (\v P \v x)_{\JJ'} /\vs p {\JJ'} \le \lambda$).
  Finally, $\lambda \le m^2 \lambda^*$
  because each $\lambda_\JJ^* \le \lambda^*$.
\end{proof}
    
Take $\lambda_0 = \lambda/m^2$ (for
$\lambda$ as in the lemma) so
\(1 \le \lambda^*/\lambda_0 \le m^2\).  
Next use binary search to find an integer $j$ such that
\(2^{j} \le \lambda^*/\lambda_0 < 2^{j+1}\).
Given an arbitrary $i$, to decide whether
\(i \le j\),
solve the feasibility subproblem taking
$\lambda' = \lambda_0 2^i$ and $\epsilon' = 1/2$.
If there is an approximate solution then
$\lambda^* \le (1+\epsilon')\lambda' < \lambda_0 2^{i+1}$
and hence $i \le j$.
Otherwise the problem is infeasible so 
$\lambda^* > \lambda' = \lambda_0 2^{i}$
and hence $i > j$. 
Since there are $O(\log m)$ possible values of $i$,
the binary search takes $O(\log \log m)$ subproblems
each with $\epsilon' = 1/2$.

We have now computed $\lambda_1 = \lambda_0 2^{j}$
such that
\(1 \le \lambda^*/\lambda_1 < 2\).
Next we increase the precision.
We start the $i$th step with $\lambda_i$ such that
\(\lambda^*/\lambda_i \in [1,1+\delta_i]\) for some $\delta_i>0$,
solve the feasibility problem
taking $\lambda' = \lambda_{i}(1+\delta_i/4)$
and $\epsilon' = \delta_i/4$.
If there is an approximate solution then
\(\lambda^*/\lambda_i \in [1,(1+\delta_i/4)^2]\),
so take $\lambda_{i+1} = \lambda_i$.
Otherwise the problem is infeasible, implying
\(\lambda^*/\lambda_i \in [1+\delta_i/4,1+\delta_i]\),
so take $\lambda_{i+1} = \lambda_i(1+\delta_i/4)$.
In either case a calculation shows that
\(\lambda^*/\lambda_{i+1} \in [1,1+\delta_{i+1}]\) 
for $\delta_{i+1}=(3/4)\delta_i$.
Before $O(\log 1/\epsilon)$ steps,
$\delta_i \le \epsilon$, at which point
the most recent solution produced
will be $\epsilon$-optimal.

\begin{figure*}[t]\figsize
  \begin{center}
    \parbox{0.90\textwidth}{
      {\bf in:} weighted, capacitated digraph $G$,
      commodities, $\{(s_i,t_i,d_i)\}$, budget $W$.
      \\{\bf out:} 'infeasible' or $f$ s.t.\ $f(e) \le \mu_e(1+O(\epsilon)),
      f(s_i,t_i)\ge d_i, w\cdot f \le (1+O(\epsilon))W$

      {

        0. Initialize $f(p) \leftarrow 0$ for all $p$.

        1. Let \(N \leftarrow 2\ln(m)/\epsilon\),
        where $m=1+\#\mbox{edges}+\#\mbox{commodities}$.

        2. Define \(\local_p(f) = 
        [w(p)e^{w\cdot f/W}/W + \sum_{e\in p} e^{f(e)/\mu_e}/\mu_e]
        / [e^{-f(s_i,t_i)/d_i}/d_i]\).

        3. Define \(\Global(f) = 
        [e^{w\cdot f/W} + \sum_e e^{f(e)/\mu_e}] / [\sum_i e^{-f(s_i,t_i)/d_i}]\).

        4. Until each commodity's demand is exceeded by a factor of
        $N$ do:

        5a.\hspace{-0.5em}\tab If $g$ is not yet set or
        \(\min_p \local_p(f)/g > 1+\epsilon\), then

        5b.\hspace{-0.5em}\tab\tab let $g \leftarrow
        \Global(f)$, and

        5c.\hspace{-0.5em}\tab\tab if $\min_p \local_p(f)/g > 1$ then return
        'infeasible'.

        6.\tab Delete any commodity
        whose demand is exceeded by a factor of $N$.

        7.\tab Choose any commodity $i$ and path $p$ for
        it s.t.\  \(\local_p(f)/g \le 1+\epsilon\).

        8.\tab Set $f(p)\leftarrow f(p)+ \delta$, where
        $\delta = \epsilon \min \{d_i, W/w(p), \min_{e\in p} \mu_e\}$.

        9. Return $f/N$.
        }
      }
    \caption{Algorithm with phases, applied to min-cost concurrent multicommodity flow.
      Implementable in time bounded by time to solve
      $O(m \log(m)/\epsilon^2)$ shortest-path subproblems, where $m$
      is the number of edges plus the number of commodities.
      Note: $f(s_i,t_i)$ denotes the flow shipped for commodity $i$.
      }
    \label{mcflow}
  \end{center}
  \figspace
\end{figure*}

The second phase (increasing the precision) requires solving
$O(\log 1/\epsilon)$ subproblems, but, because $\epsilon'$
decreases geometrically in each step, the time to solve the
subproblems is dominated by the time to solve the final
subproblem (with $\delta = \Omega(\epsilon)$).  Thus, the entire
computation time is $O(\log\log m)$ times the time to solve a
feasibility problem with $\epsilon' = 1/2$ plus the time to
solve a single feasibility problem with $\epsilon' =
\epsilon$.

\Section{Examples}

\SubSection{Min-Cost Concurrent Multicommodity Flow}

This section illustrates how to handle problems with
exponentially many variables.

An instance of the min-cost concurrent multicommodity flow
problem is defined by a weighted, capacitated, directed graph
$G$, a collection of {\em commodities} $C_1,C_2,...,C_k$, and
a demand $d_i\ge 0$ for each commodity.  Each commodity $C_i$
is the set of paths from some source vertex $s_i$ to a sink
vertex $t_i$.
We also assume we are given a budget $W\ge 0$.

A solution is a multicommodity flow $f$,
consisting of a network flow $f_i$ for each commodity $C_i$.
We think of $f_i$ as specifying a flow $f_i(p)\ge 0$ for each path $p\in C_i$,
but it also induces a flow $f_i(e) = \sum_{p\ni e} f_i(p)$ on each edge $e$.
Without loss of generality, no path is in two commodities,
so we drop the subscript $i$ from $f_i(p)$.

Let $w_e$ and $\mu_e$ denote the weight and capacity of
edge $e$, respectively.  Define the weight of path $p$ to
be $w_p =\sum_{e\in p} w_e$, and the weight of flow $f$ to be
$w\cdot f = \sum_p w_p f(p)$.  For the total flow on an edge
$e$ or path $p$, we use $f(e) = \sum_{p\ni e} f(p)$.
The amount of commodity $i$ shipped is $\sum_{p \in C_i} f(p)$.

A solution $f$ is {\em feasible} if: the amount of 
commodity $i$ shipped is at least $d_i$,
the flow on each edge is within the capacity
($f(e)\le \mu_e$),
and the weight of the flow is within the
budget
($w\cdot f \le W$).  An approximate solution, given $\epsilon>0$, is one
where $|f_i| \ge (1-\epsilon) d_i$, 
$f(e) \le (1+\epsilon)\mu_e$, and $w \cdot f \le (1+\epsilon) W$.

As described, the problem is naturally a mixed
packing/covering problem with a variable $f(p)$ 
for each path $p$ and with the following constraints:
$(\forall i) \sum_{p\in C_i} f(p) \ge d_i$,
$(\forall e) \sum_{p\ni e} f(p) \le \mu_e$,
$\sum_p f(p) w_p \le W$.

\begin{figure*}[t]\figsize
  \begin{center}
    \parbox{0.90\textwidth}{

      {\bf in:} $\v A, \epsilon$
      \\{\bf out:} 'infeasible' or $\v x$ 
      s.t.\ $1 \le \v A \v x \le 1+O(\epsilon)$.

      {
        0. Let $\vs x \II = \min_\JJ 1/(n\vs A {\JJ\II})$,
        where $n$ is the number of variables.
        
        1. Let \(N \leftarrow (1 + 2\ln m)/\epsilon\),
        where $m$ is the number of constraints.

        2. Define \(\local_\II(\v x) = 
        \sum_\JJ \vs A {\JJ\II} e^{(\v A\v x)_\JJ}
        / \sum_\JJ \vs A {\JJ\II} e^{-(\v A\v x)_\JJ}\).

        3. Define \(\Global(\v x) = 
        \sum_\JJ e^{(\v A\v x)_\JJ} / \sum_\JJ e^{-(\v A\v x)_\JJ}\).

        4. While $\min \v A \v x < N$ do:

        5a.\hspace{-0.5em}\tab If $g$ is not yet set or
        \(\min_\II \local_\II(\v x)/g > 1+\epsilon\), then

        5b.\hspace{-0.5em}\tab\tab let $g \leftarrow  \Global(\v x)$, and

        5c.\hspace{-0.5em}\tab\tab if $\min_\II \local_\II(\v x)/g > 1$ then return
        'infeasible'.

        6.\tab Delete the $\JJ$th row of $\v A$
        for each $\JJ$ s.t.\ $(\v A \v x)_\JJ \ge N$.
        \rem{unnecessary, as $\v A\v x \le O(N)$}

        7a.\hspace{-0.5em}\tab Compute $\dx$ and then
        $\delta$ such that

        7b.\hspace{-0.5em}\tab\tab
        $\dx_\II = \vs x \II$
        if \(\local_\II(\v x)/g \le 1+\epsilon\) and $\dx_\II=0$ otherwise,

        7b.\hspace{-0.5em}\tab\tab 
        and $\delta = \max_\JJ (\v A \dx)_\JJ$.

        8.\tab Set $\v x \leftarrow \v x + \epsilon \,\dx/\delta$.

        9. Return $\v x/N$.
        }
      }
    \caption{Parallel algorithm as it specializes to approximately solve $\v A \v x = \v b$
      (in normalized form $\v A\v x = 1$).
      Used in the x-ray tomography example.
      Runs in $O(\log(m)/\epsilon^2)$ phases,
      makes $O(m\log(m)/\epsilon^2)$ increments,
      and runs in time $1/\epsilon^4$ times polylogarithmic in
      $n$ and $m$.
      }
    \label{invert}
  \end{center}
  \figspace
\end{figure*}

The simple implementation of the algorithm with phases reduces
in this case to the algorithm in Fig.~\ref{mcflow}.
As presented in the figure, the algorithm uses
exponentially many variables (one for each path).  However, to
implement the algorithm it suffices to maintain only the flow
for each edge and commodity and the total cost.
To implement the inner loop,
do the following for each commodity $i$:
repeatedly find the shortest path $p$ from $s_i$ to $t_i$ 
in the graph with edge weights given by
$\ell(e) = w_e e^{w\cdot f/W} + e^{f(e)/\mu_e}/\mu_e$.
If the length of $p$ is at most $(1+\epsilon)g\, e^{-f(s_i,t_i)/d_i}/d_i$,
then $\local_p(f)/g \le (1+\epsilon)$, so 
augment flow on $p$ as described in the figure,
otherwise, move on to the next commodity.

The time the algorithm takes is bounded by the shortest path
computations.  The number of these is equal to the number of
increments plus at most one per commodity per phase.
The number of increments is $O(m \log(m)/\epsilon^2)$,
while there are  $O(\log(m)/\epsilon^2)$ phases
and $O(m)$ commodities.
Thus there are $O(m \log(m)/\epsilon^2)$ shortest path computations.

\SubSection{X-Ray Tomography / Linear Equations}

Computer tomography (a.k.a.\ x-ray tomography or Radon
transform) is a special case of mixed packing and covering.
Briefly, x-rays are taken of an object from many directions,
and the internal structure (density at each point) of the object
is reconstructed from the results.

For illustration, consider the following simple case.
Assume an object resides within an $n\times n\times n$ cube.
Discretize the cube by partitioning it into
$n^3$ $1\times 1\times 1$ subcubes in the obvious way.
Enumerate the subcubes in some order and 
introduce a variable $\vs x \II$ representing the density
of the $\II$th subcube.

Take $d$ x-ray snapshots of the object from different
directions.  With current techniques, $n$ is typically a few
hundred and $d= \Theta(n^2)$ so that enough information is
gathered to reconstruct a single solution.  
Even two-dimensional reconstruction problems are useful, as a
volume can be reconstructed in slices.

Assume each x-ray produces an $n\times n$ image.  Discretize
the image into its $n^2$ squares in the obvious way.
Enumerate all $d n^2$ squares of all snapshots in some order.
For the $\JJ$th square, compute from the darkness of the square
the total mass $\vs \mu \JJ$ of the matter that the x-rays
aimed at that square passed through.  Add a constraint of the
form $\sum_\II \vs x \II \vs A {\JJ\II} = 1$ where
$\vs A {\JJ\II}$ is the volume of the intersection of cube $\II$ and the 
cylinder of x-rays aimed at square $\JJ$,
divided by $\vs \mu \JJ$.
(If $\vs \mu \JJ = 0$, delete all variables $\vs x \II$
such that the $\vs A {ij} > 0$.)

The reconstruction problem is to find
$\v x \ge 0$ such that $\v A \v x = 1$.
The approximate version is to find $\v x$
such that
$\v A \v x \ge 1$
and $\v A \v x \le \v 1+O(\epsilon)$.
The problem has $O(d n^2)$ constraints,
and each variable occurs in $d$ constraints.

Variables for cubes known to be outside the object
can be deleted.
If appropriate, additional constraints
such as $\vs x \II \le 1$ 
(to constrain the maximum density)
can be added.

Without these additional constraints,
the problem is a special case of approximately
solving a system of equations with non-negative coefficients:
given $\v A$, finding $\v x\ge 0$ such that $\v A\v x = 1$.
The parallel algorithm, as it specializes for 
this problem, is shown in Fig.~\ref{invert}.
Note that deletion of satisfied covering constraints can be omitted
and the analysis of the algorithm will still hold,
because the packing constraints ensure that
no covering constraint exceeds its upper bound
by more than an $O(1)$ factor.

The total work done by this algorithm is more than with
traditional methods for x-ray tomography (filtered
back-projection and Fourier reconstruction).  However, this
method may be easier to parallelize.  It is more flexible, in
that additional constraints can be added.  In some cases, for
example when directions from which the snapshots can be taken
are constrained, traditional methods suffer from
ill-conditioning, whereas this approach may not.

\Section{Final Remarks}

Open problem: find an efficient width-independent
Lagrangian-relaxation algorithm for the abstract mixed-packing
covering problem:
\begin{quote}
  Find $\v x : \v P \v x \le (1+\epsilon)\v p,
  \v C \v x \ge (1-\epsilon)\v c, \v x \in \cal P$
\end{quote}
where $\cal P$ is a polytope
that can be queried by
an optimization oracle
(given $\v c$, return $\v x\in\cal P$ minimizing $\v c \cdot \v x$)
or some other suitable oracle.
The min-cost multicommodity flow example earlier in the paper
is a special case.  Although that example illustrates how to deal with
exponentially many variables, the polytope in that example
is the degenerate one $\{\v x : \v x \ge 0\}$.
A polytope that illustrates the difficulty of the general
case is ${\cal P} = \{\v x : \sum_\II \vs x \II = 1\}$.
The difficulty seems to be using variable-size increments
with {\em three} constraints: the packing constraints, the covering
constraints, and the constraint of staying in the polytope.

Find a parallel algorithm whose number of iterations
is polylogarithmic in the number of constraints,
even if the number of variables is exponential.
Find a parallel algorithm whose running time
has an $\epsilon^{-2}$ or $\epsilon^{-3}$ term instead of
the $\epsilon^{-4}$ term.

The algorithms in this paper handle any pure packing or
covering problem as a special case.  In this case the
algorithms simplify somewhat, so that they can handle the
optimization versions of the problems directly.

The algorithms can be viewed as derandomizations
(using the method of conditional probabilities)
of natural randomized rounding schemes
(see \cite{Young95,Young00} for this approach).
Lower bounds on the number of iterations
required by Lagrangian-relaxation algorithms
are presented in \cite{KleinY99}.

Thanks to Lisa Fleischer for useful discussions.
{
\bibliographystyle{ieee_cs_latex/latex8}
\bibliography{/home/neal/library/bibliography/matias/names,bib,proceedings}
}

\end{document}